\documentclass{aa}  

\usepackage{graphicx}
\usepackage{txfonts}
\usepackage{lipsum}
\usepackage{subcaption}         
\usepackage{lscape}             
\usepackage{placeins}           
                                
\usepackage{CJKutf8}
\usepackage{multirow}
\usepackage{tikz}

\usepackage[colorlinks=true, citecolor=blue, urlcolor=blue, linkcolor=blue]{hyperref} 

\begin{document}

   \title{Identifying Compton-thick active galactic nuclei in the COSMOS}
  
  \subtitle{II. Searching among mid-infrared selected AGNs}


   \author{Xiaotong Guo \begin{CJK}{UTF8}{gkai}(郭晓通)\end{CJK}
        \inst{1,2,4}
        \and
        Qiusheng Gu \begin{CJK*}{UTF8}{gkai}(顾秋生)\end{CJK*}\inst{3,4}\fnmsep\thanks{qsgu@nju.edu.cn}
        \and 
        Guanwen Fang \begin{CJK*}{UTF8}{gkai}(方官文)\end{CJK*}\inst{1,2}
        \and
        Shiying Lu \begin{CJK*}{UTF8}{gkai}(陆诗莹)\end{CJK*}\inst{1,2,4}
        \and
        Fen Lyu \begin{CJK*}{UTF8}{gkai}(吕芬)\end{CJK*}\inst{1,2}
        \and
        Yongyun Chen \begin{CJK*}{UTF8}{gkai}(陈永云)\end{CJK*}\inst{5}
        \and
        Nan Ding \begin{CJK*}{UTF8}{gkai}(丁楠)\end{CJK*}\inst{6}
        \and
        Mengfei Zhang \begin{CJK*}{UTF8}{gkai}(张孟飞)\end{CJK*}\inst{1,2}
        \and
        Xiaoling Yu \begin{CJK*}{UTF8}{gkai}(俞效龄)\end{CJK*}\inst{5}
        \and
        Hongtao Wang \begin{CJK*}{UTF8}{gkai}(王洪涛)\end{CJK*}\inst{7}
  }

   \institute{School of Physics and Astronomy, Anqing Normal University, Anqing 246133, China
        \and
        Institute of Astronomy and Astrophysics, Anqing Normal University, Anqing 246133, China\\
        \email{guoxiaotong@aqnu.edu.cn}
        \and
        School of Astronomy and Space Science, Nanjing University, Nanjing, Jiangsu 210093, China\\
        \email{qsgu@nju.edu.cn}
        \and
        Key Laboratory of Modern Astronomy and Astrophysics (Nanjing University), Ministry of Education, Nanjing 210093, China
        \and
        College of Physics and Electronic Engineering, Qujing Normal University, Qujing 655011, China
        \and
        School of Physical Science and Technology, Kunming University, Kunming 650214, China
        \and
        School of Science, Langfang Normal University,Langfang 065000, China
  }

   \date{Received September 30, 20XX}

 
  \abstract
   {Compton-thick active galactic nuclei (CT-AGNs), defined by a column density of $\mathrm{N_H} \geqslant 1.5 \times 10^{24} \ \mathrm{cm}^{-2}$, are so heavily absorbed that their X-ray emission is often feeble and can even be undetectable by X-ray instruments in some cases.
        Nevertheless, their radiation is expected to be a substantial contributor to the cosmic X-ray background (CXB), predicting that CT-AGNs would comprise at least $\sim$30\% of the total AGN population.
   }
   {Cosmological Evolution Survey (COSMOS) reported that the identified CT-AGN fraction falls far below theoretical expectations, indicating that a substantial population of CT-AGNs is hidden due to their low photon counts or due to their flux lying below the current flux limits of X-ray instruments.
        This work focuses on identifying CT-AGNs hidden among mid-infrared (MIR) selected AGNs.} 
   {
        First, we selected a sample of 1,104 MIR-selected AGNs that were covered, but individually undetected by X-ray. Next, we reduced the X-ray data in the COSMOS and analyzed multiwavelength data in our sample to derive the key physical parameters required for the CT-AGN identification.
   }
   {
        Using MIR diagnostics, we found 7 to 23 CT-AGN candidates. Their subsequent X-ray stacking analysis revealed a clear detection at $>3\sigma$ significance in the soft band and only a $>1\sigma$ significance in the hard band. We fit the stacked soft- and hard-band fluxes with a physical model and confirm that these sources are absorbed by Compton-thick material.
        However, CT-AGNs ultimately constituted only 2.1\% (23/1104) of our sample, significantly below the fraction predicted by CXB synthesis models. This indicates that a considerable population of CT-AGNs remains missed by our selection.
        A comparison of host-galaxy properties between CT-AGNs and non-CT-AGNs reveals no significant differences.
   }
   {}


   \maketitle

\section{Introduction}
Active galactic nuclei (AGNs) are defined as galaxies hosting supermassive black holes (SMBHs, $10^6\mathtt{-}10^9 \mathrm{M}_\odot$) that have an Eddington ratio\footnote{The Eddington ratio is defined as the ratio of the bolometric luminosity of the AGN to the Eddington luminosity, which represents the balance between the gravitational force and the radiation pressure. For a supermassive black hole with mass $\mathrm{M}_{BH}$, the Eddington luminosity can be approximated as $$\mathrm{L}_{Edd}= 1.5\times 10^{38}\ \frac{\mathrm{M}_{BH}}{\mathrm{M}_\odot}\ \mathrm{erg\ s}^{-1}. $$} exceeding the limit of $10^{-5}$ \citep{2015ARA&A..53..365N}.
These systems are powered by the accretion of surrounding matter through the SMBHs, thereby releasing their gravitational potential energy.
Their emissions cover the entire electromagnetic spectrum, exhibiting distinct radiation characteristics at varying distances from the central SMBHs. The AGN unification model suggests that the radiation across different bands originates from their specific substructures \citep{1993ARA&A..31..473A, 1995PASP..107..803U, 2015ARA&A..53..365N}. For instance, the accretion disk near the SMBH is heated by the viscous dissipation of gravitational potential energy during mass accretion, producing a broad ultraviolet (UV) to optical continuum spectrum \citep[e.g.,][]{1980ApJ...235..361R}. The hot corona, hovering above the accretion disk, generates luminous X-ray emission through inverse Compton up-scattering of UV–optical photons originating from the accretion disk \citep{1979ApJ...234.1105L,1991ApJ...380L..51H}. The dusty torus, situated outside the corona and accretion disk, absorbs the intense radiation from the central engine and reprocesses it into thermal emission at longer mid-infrared (MIR) band \citep[e.g.,][]{1993ARA&A..31..473A}.
Owing to the evolutionary stages of AGNs and the viewing angle effects, the majority of AGNs are obscured by large amounts of dust and gas \citep[e.g.,][]{2018ARA&A..56..625H}. The high penetration ability of X-ray radiation, combined with the bright X-ray emissions from AGNs, makes X-rays an essential tool for distinguishing AGNs from other extragalactic sources and for investigating the obscured material within AGNs \citep[e.g.,][]{2017NewAR..79...59X}. The neutral hydrogen column density ($N_\mathrm{H}$) is a key parameter for describing the obscured AGNs in the X-ray band. The obscured AGNs exhibiting a column density of $N_\mathrm{H} \geqslant  1.5\times 10^{24}\ \mathrm{cm}^{-2}$ are classified as Compton-thick AGNs \citep[CT-AGNs; e.g.,][]{2004ASSL..308..245C}.

The cosmic X-ray background (CXB) is the radiation from the superposition of all discrete X-ray point sources and AGNs account for the vast majority of these sources \citep{2000Natur.404..459M,2003ApJ...588..696M,2017ApJS..228....2L}. The most prominent observational feature of the CXB spectrum is a pronounced hump, marking the peak of its energy density, around 30~keV \citep{2003A&A...411..329R,2007A&A...467..529C,2007ApJ...666...86F,2008ApJ...689..666A}. Over the past several decades, theoretical modeling of the CXB spectrum has repeatedly demonstrated that only AGN population synthesis models incorporating a substantial fraction of CT-AGNs can successfully reproduce the complete spectral shape of the CXB \citep[e.g.,][]{2007A&A...463...79G, 2009ApJ...696..110T, 2011ApJ...728...58B, 2012A&A...546A..98A, 2014ApJ...786..104U, 2019ApJ...871..240A, 2025A&A...695A.128G}.
To date, the fraction of CT-AGNs remains uncertain and is a subject of hot debate.
At low redshifts, as in the case of the Local Universe, AGN population synthesis models predict that CT-AGNs account for 20--30\% of the total AGN population detected by hard X-ray surveys such as \textit{Swift}-BAT \citep{2011ApJ...728...58B,2014ApJ...786..104U,2015ApJ...815L..13R}. 
The denser atomic and molecular gas content in high-redshift galaxies \citep[e.g.,][]{2013ARA&A..51..105C}, relative to the Local Universe, increases the likelihood that AGNs hosted in these environments are heavily obscured, thereby leading to a higher expected fraction of CT-AGNs. For instance, \cite{2015ApJ...802...89B} estimated that CT-AGNs account for $38_{-7}^{+8}\%$ of the total AGN population in X-ray deep-field surveys, while \cite{2019ApJ...871..240A} found that this fraction increases to $56\%\pm 7\%$ for AGNs throughout redshift $z \simeq 1$.

To observationally recover the intrinsic fraction of CT-AGNs in the total AGN population, it is essential to select an AGN sample that is unbiased against CT-AGNs. AGN samples selected through MIR methods meet this criterion due to the lower extinction at this waveband \citep[e.g.,][]{2001ApJ...554..778L}. Some studies have estimated that the fraction of CT-AGNs in MIR-selected AGN samples aligns with the theoretical predictions for the CXB \citep[e.g.,][]{2023ApJ...950..127C, 2024A&A...692A.250A,2024ApJ...966..229L}. Recently, \cite{2025MNRAS.540.3827A} directly measured a CT-AGN fraction of $32^{+30}_{-18}\%$ in a MIR-selected AGN sample within 15 Mpc, where the column densities are derived from broad-band X-ray spectral analysis. Similarly, \cite{2025ApJ...978..118B} directly measured the CT-AGN fraction of $35\%\pm 9\%$ in a sample of 122 nearby ($z < 0.044$) AGNs primarily selected for their MIR colors. 
These results have already provided compelling evidence to support the AGN synthesis models of the CXB.

Nevertheless, attaining the theoretically anticipated fraction of CT-AGNs within X-ray surveys remains a significant challenge.
The X-ray radiation of CT-AGNs is typically extremely faint, so their photon counts are low and often even undetectable. These characteristics make CT-AGNs the most difficult population to detect in X-ray surveys \citep[e.g.,][]{2000MNRAS.318..173M,2015A&A...574L..10C,2017A&ARv..25....2P}. In particular, for sources with column densities $N_\mathrm{H}\geqslant 10^{25}\ \mathrm{cm}^{-2}$, even in the Local Universe, X-ray flux-limited surveys are biased against detecting CT-AGNs \citep[e.g.,][]{2015ApJ...815L..13R,2016ApJ...825...85K,2025MNRAS.540.3827A,2025ApJ...978..118B}.
X-ray surveys typically report an observed fraction of CT-AGNs below 15\% \citep[e.g.,][]{2017ApJS..233...17R,2018ApJS..235...17M,2019A&A...621A..28G}, which is significantly lower than theoretical model predictions, meaning that a large number of CT-AGNs remain unidentified or undiscovered. These missing CT-AGNs may still be hidden in AGNs that exhibit low photon counts \citep[e.g.,][]{2020ApJ...897..160L} or that have not been detected by X-ray surveys \citep[e.g.,][]{2021ApJ...908..185C}. 
To identify the missing CT-AGNs in the Local Universe, the X-ray astronomy group at Clemson University and the extragalactic astronomy group at the INAF-OAS Bologna and INAF-IASF Palermo initiated a targeted project in 2017 to actively search for CT-AGNs, known as the Clemson-INAF CT-AGN Project \citep{2018ApJ...854...49M}. 
The project used a volume-limited sample of CT-AGNs with NuSTAR spectra, achieving a CT-AGN fraction of $20\%\pm 5\%$ within the redshift of $z<0.01$ \citep[][]{2021ApJ...922..252T}. Similarly, a series of studies have focused on searching for the missing CT-AGNs in X-ray deep-field surveys. For example, \cite{2021ApJ...908..169G} identified eight missing CT-AGNs by using the multiwavelength techniques in the \textit{Chandra} Deep Field-South (CDFS) survey. \cite{2025ApJ...987...46Z} used machine learning algorithms to identify 67 missing CT-AGNs, thereby raising the fraction of CT-AGNs to 20\% in the CDFS survey. Numerous efforts have been made to search for the missing CT-AGNs in X-ray surveys, yet the theoretical expectations remain unmet.

The Cosmic Evolution Survey \citep[COSMOS; e.g.,][]{2007ApJS..172....1S} is a deep, multiwavelength survey covering a relatively large area of 2 deg$^2$, aimed at measuring the evolution of galaxies and AGNs. The X-ray observations of this field are also crucial for resolving the CXB \citep[e.g.,][]{2015ApJ...802...89B}. The \textit{Chandra} X-ray telescope has covered the entire COSMOS, achieving a relatively deep average exposure \citep[e.g.,][]{2016ApJ...819...62C,2016ApJ...817...34M}. However, the observed fraction of CT-AGNs in the COSMOS is significantly lower than theoretical expectations. For example, \cite{2018MNRAS.480.2578L} successfully identified 67 CT-AGNs from a sample of 1855 AGNs with high (>30) photon counts through X-ray spectral fitting. To find out the hidden CT-AGNs in the COSMOS, it is essential to establish and execute a series of specific tasks. \cite{2025A&A...694A.241G}  completed the first task and successfully identified 18 CT-AGNs. In this work, we aim to execute and complete the second task, focusing on identifying CT-AGNs among MIR-selected AGNs.
This paper is structured as follows.
Section~\ref{sec:sample} presents the data used to select AGNs via MIR colors and details the construction of our final sample.
In Sect.~\ref{sec:x-ray-data} we describe the \textit{Chandra} observations covering COSMOS and their data reduction.
Section~\ref{sec:mult-data} presents the multiwavelength photometric data for the sample and the corresponding analysis.
Section~\ref{sec:diagnostic} presents our CT-AGN diagnostics, along with a brief discussion of the missing CT-AGNs in our sample and the properties of CT-AGN host galaxies.
A brief summary is presented in Sect.~\ref{sec:summary}.
We adopted a concordance flat $\Lambda$-cosmology with ${\rm H_0 = 67.4 \ km\ s^{-1}\ Mpc^{-1}}$, $\Omega_{\rm m} = 0.315$, and $\Omega_\Lambda = 0.685$ \citep{2020A&A...641A...6P}.
All quoted uncertainties correspond to the 1$\sigma$  (68\%) confidence level unless stated otherwise, while upper limits are given at the 3$\sigma$ level.

\section{The MIR-selected AGN sample}\label{sec:sample}
\subsection{MIR data collation}

To construct a high-quality MIR-selected AGN sample, we systematically collated the MIR data. The primary data source was the COSMOS2020 FARMER catalog \citep{2022ApJS..258...11W}, which provides \textit{Spitzer}/IRAC four-band (3.6~$\mu$m, 4.5~$\mu$m, 5.8~$\mu$m, and 8.0~$\mu$m) photometric data. To meet the requirements for the sample selection, the following constraints were applied to the MIR photometric data:
\begin{enumerate}[(1)]
        \item Sources with redshifts between 0.1 and 2 were selected;
        \item Sources with IRAC magnitudes over the $3\sigma$ detection limit were chosen to ensure data reliability;
        \item The flux-to-error ratio in each band was required to be greater than 5 to ensure a high signal-to-noise ratios.
\end{enumerate}
In the COSMOS2020 FARMER catalog, a total of 56,831 sources satisfy the above constraints. We  collected the four band flux densities of \textit{Spitzer}/IRAC for the steps described in the next section.

\subsection{Selection of AGN sample}
\begin{figure}
        \includegraphics[width=\columnwidth]{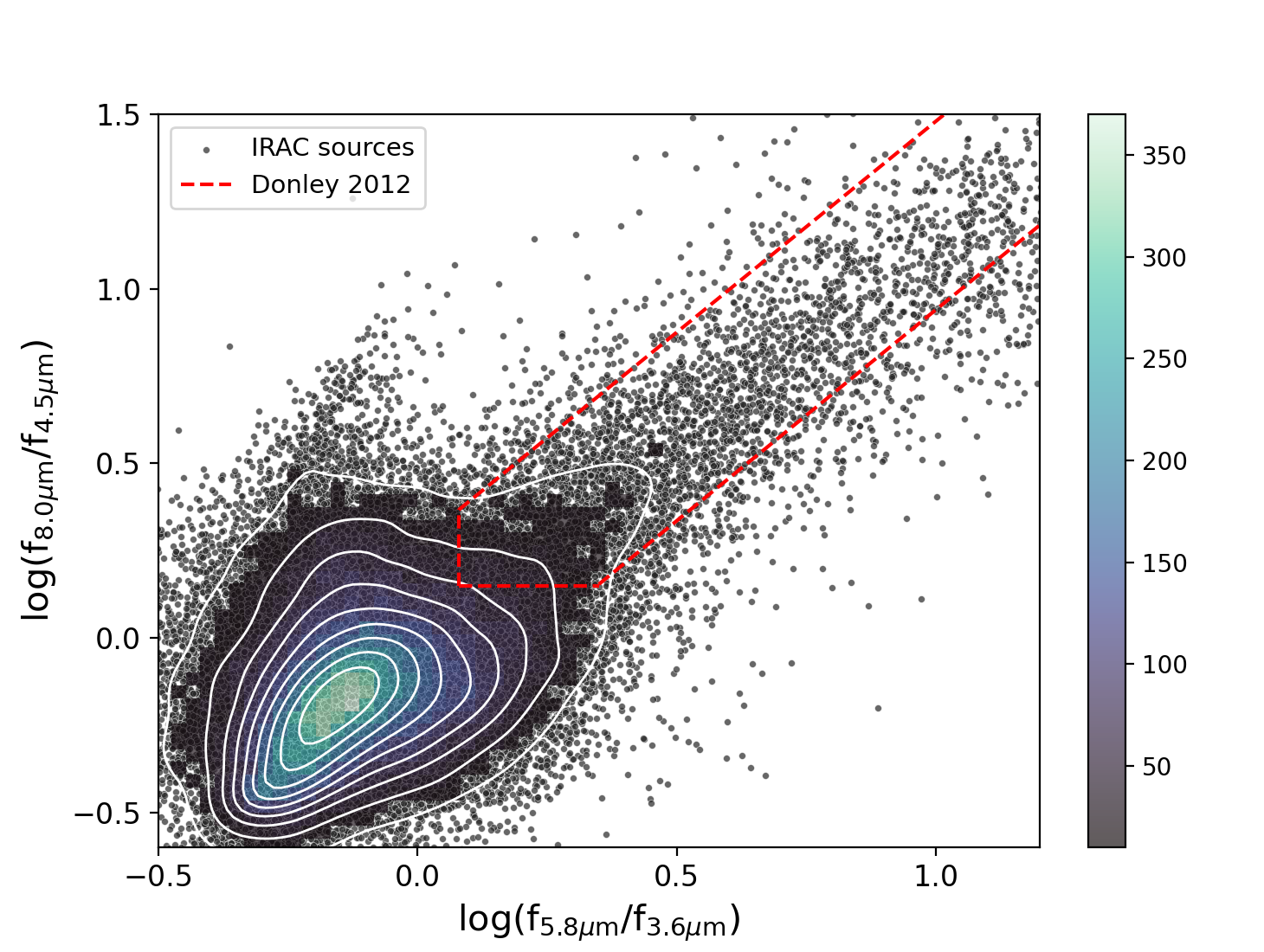}
        \caption{IRAC color–color diagram of the MIR sources. The small gray dots represent all IRAC sources selected in the COSMOS. The region bounded by dashed red lines marks the MIR-AGN selection region of \cite{2012ApJ...748..142D}. The sources within this region are classified as MIR-selected AGNs. The color bar and overlaid contours trace the number density gradient across the diagram.}
        \label{fig:MIR-AGN}
\end{figure}

The unique colors of AGNs in the MIR band, especially between 3--5 $\mu$m, make their spectra appear redder than those of normal galaxies. Therefore, many studies use this feature to distinguish AGNs from normal galaxies and identify them \citep[e.g.,][]{2004ApJS..154..166L,2007AJ....133..186L,2013ApJS..208...24L,2012ApJ...748..142D,2013ApJ...772...26A,2017ApJS..233...19C}. For example, the IRAC color-color diagrams most commonly used for AGN selection were defined by \cite{2004ApJS..154..166L,2007AJ....133..186L,2013ApJS..208...24L} and \cite{2012ApJ...748..142D}. 
In this work, we adopted the AGN selection criterion of \citet{2012ApJ...748..142D}, which is more stringent than others and yields a highly pure and reliable AGN sample.
Figure~\ref{fig:MIR-AGN} presents an IRAC color-color diagram, plotting $\log(f_{5.8\mu \mathrm{m}}/f_{3.6\mu \mathrm{m}})$ against $\log(f_{8.0\mu \mathrm{m}}/f_{4.5\mu \mathrm{m}})$. The sources within the red dashed lines are AGNs selected using the criteria from \cite{2012ApJ...748..142D}, totaling 1,545. 
Among them 252 are X-ray sources, of which nine have been classified as CT-AGNs by \cite{2018MNRAS.480.2578L} and \cite{2025A&A...694A.241G}.

To identify CT-AGNs among AGNs undetected by X-ray in the COSMOS, we applied the following constraints to the MIR-selected AGN sample:
\begin{enumerate}[(1)]
        \item Remove known X-ray sources from the MIR-selected AGNs;
        \item Eliminate sources lacking X-ray observations (i.e., those with exposure time = 0);
        \item To avoid overestimating the flux upper limits (see Sect.~\ref{sec:upper}) of MIR-selected AGNs due to nearby bright X-ray sources, we exclude any sources located within $6''$ of known X-ray sources.
\end{enumerate}
After this screening, our final AGN sample comprises 1,104 sources. Columns 1--3 of Table~\ref{Tab:summary} list their IDs, RA, and DEC.
The photometric redshifts of our AGN sample are taken from the COSMOS2020 FARMER catalog, measured by \texttt{LePhare} using the galaxy templates. For the bright sources, \cite{2022ApJS..258...11W} find a normalized median absolute deviation of only 0.009 in their photometric redshifts; even in the faintest bin (25 mag < \texttt{i} < 27 mag) the normalized median absolute deviation is 0.036. Column 4 of Table~\ref{Tab:summary} lists their photometric redshifts.

\section{X-ray observations and data reduction}\label{sec:x-ray-data}
\subsection{Overview of Chandra's observations in COSMOS}

To identify (or probe) AGNs with varying redshifts in the COSMOS and investigate black hole accretion history, as well as the co-evolution of AGNs with their host galaxies, \cite{2009ApJS..184..158E} conducted \textit{Chandra} observations during 2006--2007, accumulating a total exposure of 1.8 Ms. These observations covered the central 0.5 deg$^2$ with an effective exposure of approximately 160 ks and an outer 0.4 deg$^2$ area with an effective exposure of approximately 80 ks. 
Subsequently, \cite{2016ApJ...819...62C} added further \textit{Chandra} observations during 2013–2014, increasing the total exposure to 2.8 Ms. These additional observations covered the remaining area \citep[the outer 0.4 deg$^2$ from ][]{2009ApJS..184..158E} with an effective exposure of approximately 80 ks and increased the effective exposure over the central 1.5 deg$^2$ to approximately 160 ks. As a result, the combined observations totaling approximately 4.6 Ms covered a region of 2.2 deg$^2$ in the COSMOS.
For more detailed information on observation files, including observation and exposure times, we refer to \cite{2009ApJS..184..158E}  and \cite{2016ApJ...819...62C}. 
\subsection{Data acquisition and reduction}
To determine the X-ray flux upper limits for each AGN in our sample, their exposure times and counts are required. As the exposure and image maps are not archived, we first retrieved all 117 raw event files covering the COSMOS from the \textit{Chandra} X-ray Center\footnote{\url{https://cda.harvard.edu/chaser/}}. Next, we processed these 117 raw event files through the following detailed steps to extract the required exposure times and counts.

We processed the data reduction using \texttt{CIAO 4.17} software tools \citep{2006SPIE.6270E..1VF} and \texttt{CALDB 4.12.0}. 
We used the \textit{\texttt{chandra\_repro}} script to reprocess the data, automating the \texttt{CIAO} recommended steps and generating new level 2 event files.
Subsequently, we removed \texttt{ACIS-I} Background Flares using the \textit{\texttt{dmcopy}} and \textit{\texttt{dmextract}} scripts, following the standard Science Threads from the \textit{Chandra} X-ray Observatory.
Finally, all 117 observations were merged with the \textit{\texttt{merge\_obs}} script to produce exposure-corrected images and the corresponding modeled background maps.

\subsection{X-ray luminosity upper limit estimations}\label{sec:upper}
\begin{figure}
        \includegraphics[width=1\columnwidth]{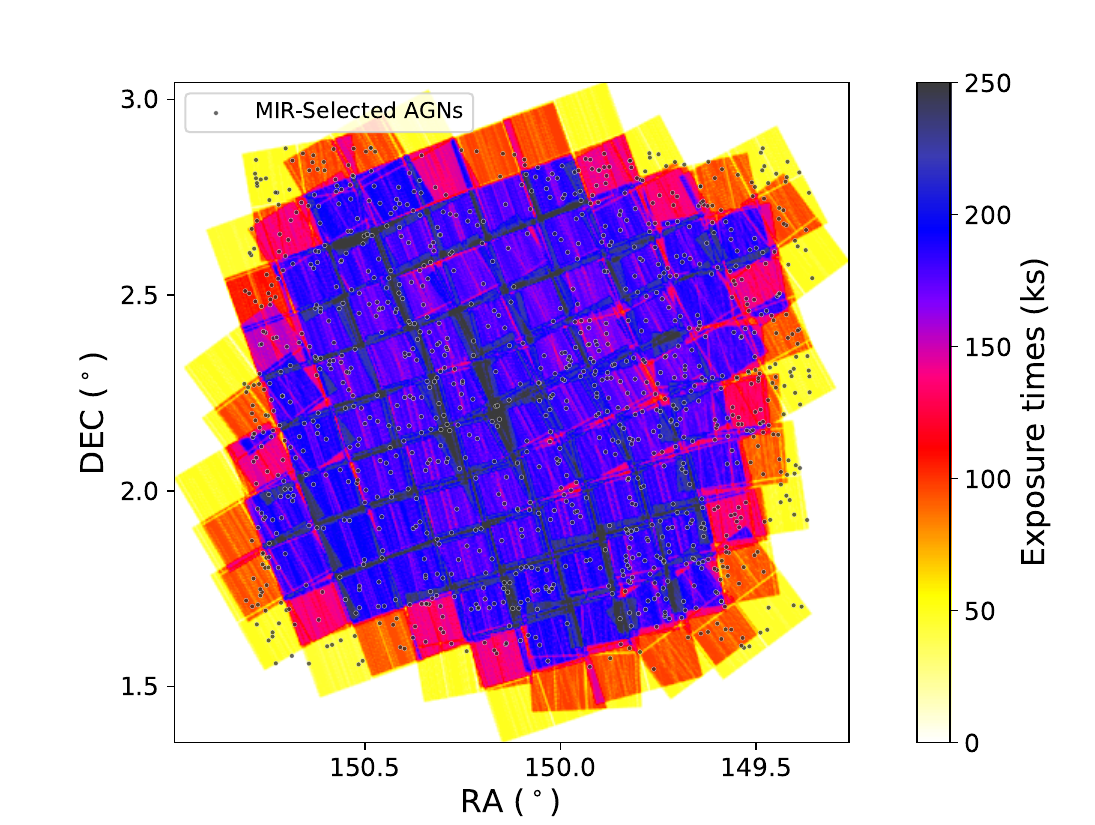}
        \caption{Merged \textit{Chandra} exposure map of the COSMOS, overlaid with the positions of MIR-selected AGNs (black dots) used in this work. The color bar indicates the effective exposure time for the corresponding X-ray coverage.}
        \label{fig:expmap}
\end{figure}

Our MIR-selected AGNs are individually undetected in X-rays, their X-ray luminosities cannot be measured directly. Consequently, we quote only upper limits, calculated with the equation \citep{2003AJ....125..383A}:\begin{equation*}
        L_{\text{2--10}}^{\text{up-limit}} = 4 \pi \cdot D_L^2 \cdot f_{\text{2--10}}^{\text{up-limit}} \cdot (1+z)^{\Gamma-2},
\end{equation*}
where $D_L$ is the luminosity distance, $f_{\text{2--10}}^{\text{up-limit}}$ represents the upper limit of the flux in the 2--10~keV within the observed frame, $z$ is the redshift, and $\Gamma$ is the photon index used for k-corrections, which is fixed to the value of 1.4 \citep{2016ApJ...819...62C}. The upper limit of the X-ray flux density is calculated by
\begin{equation*}
        f_{\text{2--10}}^{\text{up-limit}} = R_{\text{lim}} \times (\text{CF} \times 10^{-11})\ \text{erg} \cdot \text{s}^{-1},
\end{equation*}
where $R_{\text{lim}}$ is the upper limit of the count rate at the undetected source and CF is the energy conversion factor, set to 3.06 \citep{2016ApJ...819...62C}. The upper limit of the count rate for undetected sources is calculated by
\begin{equation*}
        R_{\text{lim}} = \frac{B_{3''}^{3\sigma}}{T_{\mathrm{exp}}},
\end{equation*}
where $T_{\mathrm{exp}}$ represents the effective exposure time at the source location
, $B_{3''}^{3\sigma}$ denotes the $3\sigma$ upper limit on the background counts measured within a $3''$ radius aperture centered on the source. The $3\sigma$ upper limit on the background counts can be calculated using Eq. 9 of \cite{1986ApJ...303..336G}, giving
\[B_{3''}^{3\sigma}\approx (B_{3''} +1) \left[1-\frac{1}{9(B_{3''} +1)} + \frac{3}{3\sqrt{B_{3''} +1}}\right]^3 ,\]
where $B_{3''}$ represents the 0.5--8~keV background counts measured within a $3''$-radius aperture centecalibrered on the source, after correction for the energy-encircled fraction.
The background counts for AGNs are determined by summing all values from a $3''$ radius circle centered at their respective coordinates in the modeled background map.  Similarly, the exposure times for the AGNs are determined by reading the values from the exposure map at their respective coordinates.
Figure~\ref{fig:expmap} shows the coordinates of MIR-selected AGNs on the exposure map and their corresponding exposure times. Columns 5--6 of Table~\ref{Tab:summary} give the $3''$ aperture background counts $B_{3''}$ and exposure time $T_{\mathrm{exp}}$ for each source.

Applying the calculation process described above, we derived X-ray luminosity upper limits for the entire MIR-selected AGN sample. Columns 7--8 of Table~\ref{Tab:summary} list the upper limits of flux density and luminosity in the 2--10~keV band.

\section{Analysis of multiwavelength data}\label{sec:mult-data}
\subsection{Multiwavelength photometric data}
Identifying CT-AGNs among MIR-selected AGNs requires ascertaining the MIR radiation from AGNs. Thus, we need to construct the spectral energy distributions (SEDs) for these sources to decompose the AGN's contribution to the MIR emission. This process requires multiwavelength photometric data.

The multiwavelength photometric data used in this work span the far-UV (FUV; 1526~\AA) to the far-infrared (FIR; 500~$\mu$m).
Among them, the photometric data covering the wavelength range from the FUV to the MIR (27 bands in total) are sourced from the COSMOS2020 FARMER catalog \citep{2022ApJS..258...11W}. These bands include the GALEX  FUV and NUV band, MegaCam U band, CFHT U band, five Subaru HSC bands (g, r, i, z, and y), 12 Subaru Suprime-Cam medium bands (IB427, IB464, IA484, IB505, IA527, IB574, IA624, IA679, IB709, IA738, IA767, and IB827), four VISTA  VIRCAM broad bands (Y, J, H, and $K_s$), and four \textit{Spitzer} IRAC bands (ch1, ch2, ch3, and ch4). 
Moreover, the FIR photometric data are sourced from the COSMOS2015 catalog \citep{2016ApJS..224...24L},  including MIPS 24~$\mu$m, PACS 100~$\mu$m, PACS 160~$\mu$m, SPIRE 250~$\mu$m, SPIRE 350~$\mu$m, and SPIRE 500~$\mu$m. 
Where photometric data in a given band are missing, we use the corresponding flux upper limits in the SED analysis.

\subsection{Analysis of SEDs}
\begin{figure}
        \includegraphics[width=\columnwidth]{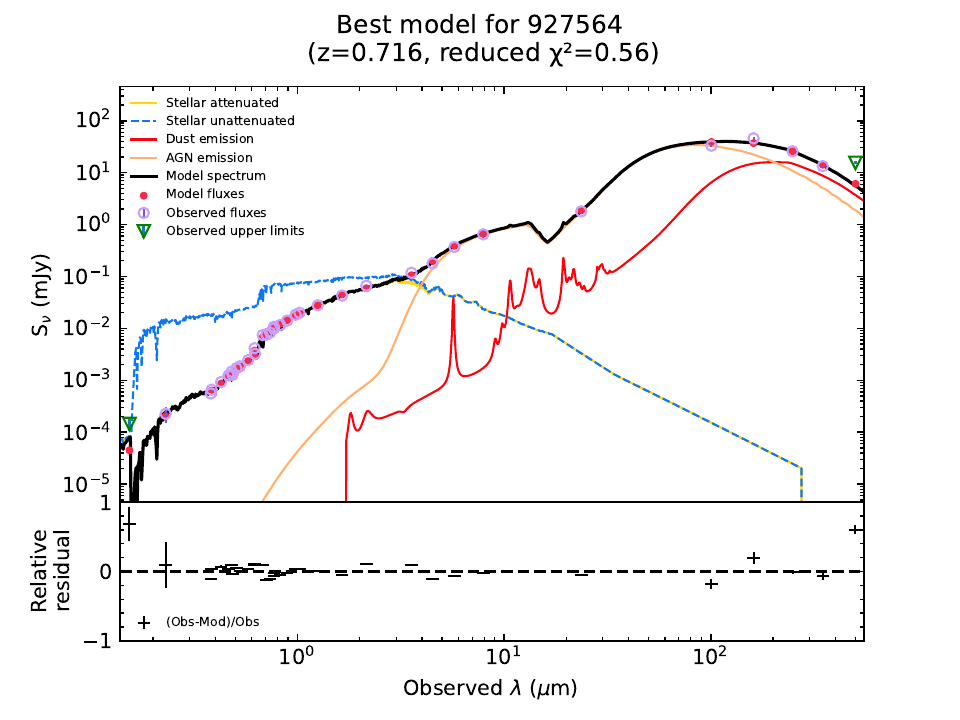}
        \caption{Example of the best-fitting SED for an AGN. The solid black line indicates the best-fitting model. The dashed blue, solid gold, and solid red lines represent unattenuated stellar, attenuated stellar, and dust emission, respectively. The solid apricot line indicates AGN emission. The strawberry filled circles, pastel-purple open circles, and green open inverted triangles denote model predictions, observed fluxes, and observed upper limits, respectively. The lower panel indicates the residual of the best fitting.}  
        \label{fig:SED-example}
\end{figure}

The SED analysis of AGNs is a widely used technique that not only provides detailed information about AGNs but also reveals the properties of their host galaxies \citep[e.g.,][]{2020MNRAS.492.1887G}, such as AGN luminosity, rest-frame 6~$\mu$m~luminosity for the AGN, stellar mass (M$_\star$), and the star formation rate (SFR). 
To conduct the analysis of AGNs in our sample and obtain the required parameters in this work, we used Code Investigating GALaxy Emission \citep[CIGALE; V2022.1;][]{2019A&A...622A.103B}, an open \textit{Python} code containing the template of galaxies and AGNs. 
Since all the sources in our sample are MIR-selected AGNs, we have used templates that incorporate both the galaxy and AGN characteristics to achieve the most accurate SED fitting.
The galaxy templates in CIGALE are generated by integrating four distinct modules, including the star formation history (SFH), the single stellar population (SSP) model \citep{2003MNRAS.344.1000B}, the dust attenuation \citep[DA;][]{2000ApJ...533..682C}, and the dust emission \citep[DE;][]{2007ApJ...657..810D,2014ApJ...784...83D}. The AGN component in our analysis is a module developed by \cite{2006MNRAS.366..767F}. The modules and parameters used for SED fitting are summarized in Table~\ref{Tab:SED-module}. 
Additionally, for sources using upper flux limits in certain bands, the \textit{lim\_flag} is set to "\texttt{noscaling}."  
Given the vast parameter space, quickly constructing a model on the computer is not feasible. Thus, we adopted an iterative methodology to expedite the acquisition of the best-fitting SEDs.
Figure~\ref{fig:SED-example} shows the best-fitting SED for an AGN. 
To assess the reliability of the fitted AGN components, we refit SEDs using only the galaxy templates. A markedly better fit (indicated by a lower $\chi^2$) with the combined galaxy+AGN templates than with the galaxy-only templates implies that the fitted AGN components can indeed represent the contribution of the actual AGNs in their SEDs.
We used the Bayesian information criterion (BIC) to quantify the significance or confidence of the AGN components \citep[see Appendix~C of][for details]{2025A&A...694A.241G}.

\begin{table*}
        \caption{Module assumptions for SED fitting.}             
        \label{Tab:SED-module}      
        \centering          
        \resizebox{\hsize}{!}
        {
                \begin{tabular}{l|c|c|c}     
                        \hline\hline       
                        Component & Module & Parameter & Value \\
                        \multicolumn{1}{c|}{(1)}&(2)&(3)&(4)\\
                        \hline                    
                        \multirow{13}{*}{Galaxy}&\multirow{5}{*}{SFH (delayed)}&tau\_main (Myr)&20 -- 8000 (in steps of 10)\\
                        &&age\_main (Myr)&200 -- 13000 (in steps of 10)\\
                        &&tau\_burst (Myr)&10 -- 200 (in steps of 1)\\
                        &&age\_burst (Myr)&10 -- 200 (in steps of 1)\\
                        &&f\_burst &0, 0.0001, 0.0005, 0.001, 0.005, 0.01, 0.05, 0.1, 0.15,0.20, 0.25, 0.3, 0.40, 0.50\\
                        \cline{2-4}
                        &\multirow{2}{*}{SSP (BC03)}&imf&1 (Chabrier)\\
                        &&metallicity &0.02\\
                        \cline{2-4}
                        &DA&\multirow{2}{*}{E\_BV\_nebular (mag)}&0.005, 0.01, 0.025, 0.05, 0.075, 0.10, 0.15,0.20, 0.25, \\
                        &(dustatt\_calzleit)&&0.30, 0.35, 0.40, 0.45, 0.50, 0.55, 0.60\\ 
                        \cline{2-4}
                        &\multirow{4}{*}{DE (dl2014)}&qpah& 1.12, 1.77, 2.50, 3.19\\
                        &&umin &5.0, 6.0, 7.0, 8.0, 10.0, 12.0, 15.0, 17.0, 20.0, 25.0\\
                        &&alpha &2.0, 2.1, 2.2, 2.3, 2.4, 2.5, 2.6, 2.7, 2.8\\
                        &&gamma &0.02\\
                        \hline
                        \multirow{8}{*}{AGNs}&\multirow{8}{*}{AGN (Fritz2006)}&r\_ratio&10, 30, 60, 100, 150\\
                        &&tau&0.1, 0.3, 0.6, 1.0, 2.0, 3.0\\
                        &&beta&-1.00, -0.75, -0.50, -0.25, 0.00\\
                        &&gamma&0.0, 2.0, 4.0, 6.0\\
                        &&opening\_angle&60, 100, 140\\
                        &&psy&0.001, 10.1, 20.1, 30.1, 40.1, 50.1, 60.1, 70.1, 80.1, 89.99\\
                        &&\multirow{2}{*}{fracAGN}&0.0, 0.05, 0.1, 0.15, 0.2, 0.25, 0.3, 0.35, 0.4, 0.45, 0.5, 0.55, 0.6, 0.65, 0.7, 0.75, \\
                        &&&0.8, 0.85, 0.9,
                        0.95, 0.99\\
                        \hline                  
                \end{tabular}
        }
\end{table*}

Through SED fitting, we obtained the best-fitting SED for each source and derived the corresponding physical parameters.
Columns 9--10 of Table~\ref{Tab:summary} are rest-frame 6~$\mu$m~luminosity for the AGN and the AGN components' confidence. Column 11 of Table~\ref{Tab:summary} presents M$_\star$, the derived outcome of SED fitting. However, since the SFR derived from SED fitting relies on the chosen SFH model \citep{2019RAA....19...39G}, we instead use the calibration method proposed by \cite{2005ApJ...625...23B} to estimate the SFR using UV and IR luminosities. This calibration is scaled to the \cite{2003PASP..115..763C} initial mass function for accuracy. Specifically, the SFR is calculated using the formula,
\begin{equation}
        \label{equ:SFR}
        \mathrm{SFR(M_\odot\ yr^{-1})}=1.09\times10^{-9} (2.2\mathrm{L_{UV}}+\mathrm{L_{IR}}),
\end{equation}
where $\mathrm{L_{UV}}=\nu \mathrm{L}_\nu$ is an estimation of the integrated 1216 -- 3000~\AA~rest-frame UV luminosity, and $\mathrm{L_{IR}}$ is the 8 -- 1000~$\mu$m~rest-frame IR luminosity. Both $\mathrm{L_{UV}}$ and $\mathrm{L_{IR}}$ are in units of $ \mathrm{L}_\odot$. Column 12 of Table~\ref{Tab:summary} lists the SFR of each source.

\section{Diagnostic CT-AGNs and discussion}\label{sec:diagnostic}

\subsection{MIR diagnostics}
The emissions from the corona and dust torus of AGNs are considered good tracers of the accretion disk's emission. Consequently, a strong correlation is expected between X-ray and MIR emissions in AGNs. Several studies have investigated this relationship in radio-quiet AGNs \citep[e.g.,][]{2009ApJ...693..447F,2015MNRAS.449.1422M,2015ApJ...807..129S,2017ApJ...837..145C,2019ApJ...886..125E,2021A&A...651A..91E}.
Due to the low optical depth in the MIR band, the MIR radiation from CT-AGNs is not significantly suppressed. 
As the column density increases, the torus absorbs more nuclear X-ray emission and re-radiates it in the MIR, potentially boosting the MIR luminosity. However, this effect is extremely weak: the AGN X-ray luminosity is 1--2 dex below its UV–optical continuum \citep{2006AJ....131.2826S, 2010A&A...512A..34L}, so even complete absorption and re-radiation in the MIR would contribute negligibly compared with the direct disk-dust heating. \cite{2017ApJS..233...17R} indeed found no significant increase in MIR luminosity with column density. Consequently, the intrinsic X-ray and MIR luminosities of AGNs are still expected to follow the above relations.
However, because the X-ray radiation of CT-AGNs is heavily absorbed, their observed X-ray luminosities fall below the values predicted by these relations.
Given that the AGNs in our sample are not detected in X-rays, we adopted the corresponding X-ray luminosity upper limits as their observed luminosities and use these relationships to identify CT-AGNs. 
\begin{figure}
        \includegraphics[width=\linewidth]{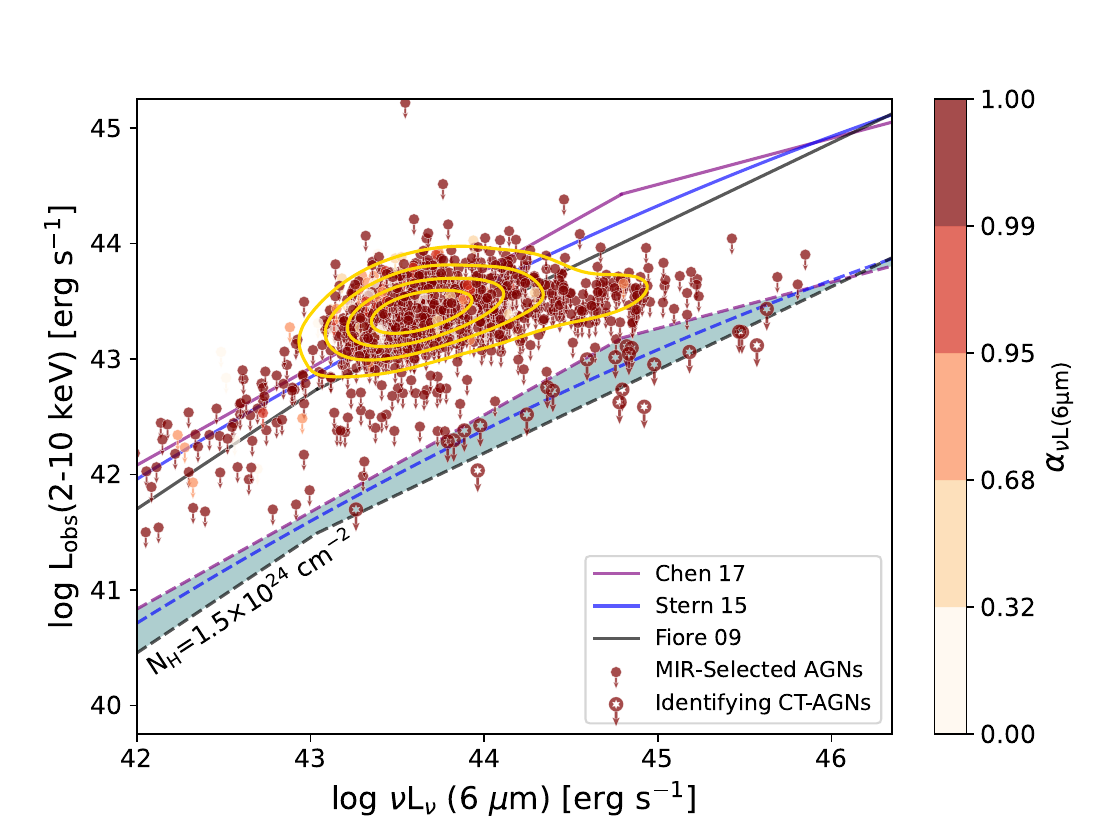}
        \caption{Observed X-ray luminosity in the rest-frame 2--10~keV band as a function of the 6~$\mu$m~luminosity for fit AGN component. The solid purple, blue, and black lines represent the relation for \cite{2017ApJ...837..145C}, \cite{2015ApJ...807..129S}, and \cite{2009ApJ...693..447F}, respectively. The dashed lines indicate the same relationships but where the X-ray luminosities are absorbed by a column density of $\mathrm{N_H}=1.5\times 10^{24}\ \mathrm{cm}^{-2}$. 
                The circles are MIR-selected AGNs with a fit AGN component. The color bar represents  the confidence coefficient of the 6~$\mu$m~luminosity for a fit AGN component (or the fit AGN components). The cutting stars within the large circles represent CT-AGNs identified using the relationship between MIR and X-ray luminosities.}
        \label{fig:midIR-diagnostics}
\end{figure}

Figure~\ref{fig:midIR-diagnostics} shows the observed X-ray luminosity in the rest-frame 2--10~keV band as a function of the 6~$\mu$m~luminosity for the fit AGN component. The solid purple, blue, and black lines represent the relation for \cite{2017ApJ...837..145C}, \cite{2015ApJ...807..129S}, and \cite{2009ApJ...693..447F}, respectively. 
Assuming the intrinsic 2--10~keV spectrum of each AGN is modeled in in \textit{Xspec} with \textit{\texttt{phabs$\ast$powerlaw+constant$\ast$powerlaw+pexmon}} (see Appendix~D of \citealt{2025A&A...694A.241G} for details), the dashed lines represent the same relationships between MIR and X-ray luminosities after the X-ray spectra have been absorbed by gas with a column density of $\mathrm{N_H}=1.5\times 10^{24}\ \mathrm{cm}^{-2}$. This means that the sources within or below the shaded area in Fig.~\ref{fig:midIR-diagnostics} should be classified as CT-AGNs. 
Circles tipped with downward arrows denote the AGNs in our sample, their colors indicating the confidence levels of the fitted AGN components.
As shown in Fig.~\ref{fig:midIR-diagnostics}, 23 AGNs fall within or below the shaded area, suggesting that they may be CT-AGNs.
However, considering the variability in the relations provided by \cite{2017ApJ...837..145C} and \cite{2009ApJ...693..447F},  a range of possible CT-AGN candidates is 7 to 23.
To further determine whether they are CT-AGNs, we must assess the reliability of their fitted AGN components. These sources, whose fitted AGN components exhibit a high confidence coefficient ($\alpha_{AGN} > 0.95$), support a CT-AGN classification; whereas those with a low confidence coefficient ($\alpha_{AGN} < 0.95$) cannot be reliably identified as CT-AGNs. 
Finally, a total of 23 AGNs located within or below the shaded region exhibit high confidence coefficients, supporting their diagnosis as CT-AGNs. These CT-AGNs are marked in Fig.~\ref{fig:midIR-diagnostics} by cutting stars within the large circles. Column 13 of Table~\ref{Tab:summary} also lists their classification, "1" represents CT-AGN identified in this work.

\subsection{Stacking analysis for CT-AGNs}
\begin{figure}
        \includegraphics[width=\linewidth,trim= 10 0 10 0]{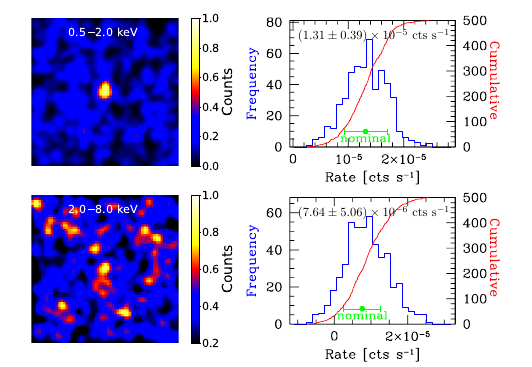}
        \caption{X-ray stacking results for the 23 CT-AGNs identified by MIR diagnostics. Left: Stacked images in the 0.5--2~keV (top) and 2--8~keV (bottom) bands. Right: Corresponding bootstrap histograms of the net count rates. Green lines indicate the mean count rates and $1\sigma$ confidence intervals. Detections exceed $3\sigma$ significance in the soft band, while only exceeding a $1\sigma$ significance in the hard band.} 
        \label{fig:stacking}
\end{figure}

In the preceding section, we successfully find out 7 to 23 CT-AGN candidates. Because these AGNs are individually undetected in X-rays, we cannot extract their X-ray spectra and determine their column densities. To confirm heavy absorption in the X-ray band, we perform X-ray stacking analysis on these 23 AGNs using the publicly available CSTACK V4.5 tool \citep{2008HEAD...10.0401M}\footnote{CSTACK was developed by Takamitsu Miyaji and is available at \url{https://lambic.astrosen.unam.mx/cstack/}}. 
This analysis adopts the default maximum off-axis angle, $3''$ source aperture, $6''$ background annulus, and automatic exclusion of known X-ray sources.
The output of CSTACK includes the stacked photon count rates and their errors for both the soft (0.5--2~keV) and hard (2--8~keV) bands, along with the corresponding stacked images.
Figure~\ref{fig:stacking} presents the stacking results: the stacked AGNs are clearly detected at $>3\sigma$ significance in the soft band and marginally detected at $>1\sigma$ in the hard band.
The stacked photon count rates in the soft and hard bands are $(1.31\pm 0.39)\times 10^{-5}\ \mathrm{cts\ s^{-1}}$ and $(7.64\pm 5.06)\times 10^{-6}\ \mathrm{cts\ s^{-1}}$, respectively. 

Assuming the AGN X-ray emission follows a simple power-law spectrum with photon flux density \(F(E)\propto E^{-\Gamma}\), we convert the soft- and hard-band count rates into photon fluxes by adopting effective areas of 600 cm$^2$ and 350 cm$^2$, and mean photon energies of 1.2~keV (soft) and 4.0~keV (hard). This derives a photon index \(\Gamma=0.0\pm 0.6\). 
 Assuming a Galactic absorption of \(N_{\rm H}=2.6\times10^{20}\ \mathrm{cm}^{-2}\), we use a tool of PIMMS V4.15\footnote{\url{https://cxc.harvard.edu/toolkit/pimms.jsp}} to estimate soft- and hard-band fluxes of \((7.57\pm 2.25) \times10^{-17}\ \mathrm{erg\ cm^{-2}\ s^{-1}}\) and \((2.34\pm 1.55)\times10^{-16}\ \mathrm{erg\ cm^{-2}\ s^{-1}}\), respectively.

\begin{figure}
        \includegraphics[width=\linewidth]{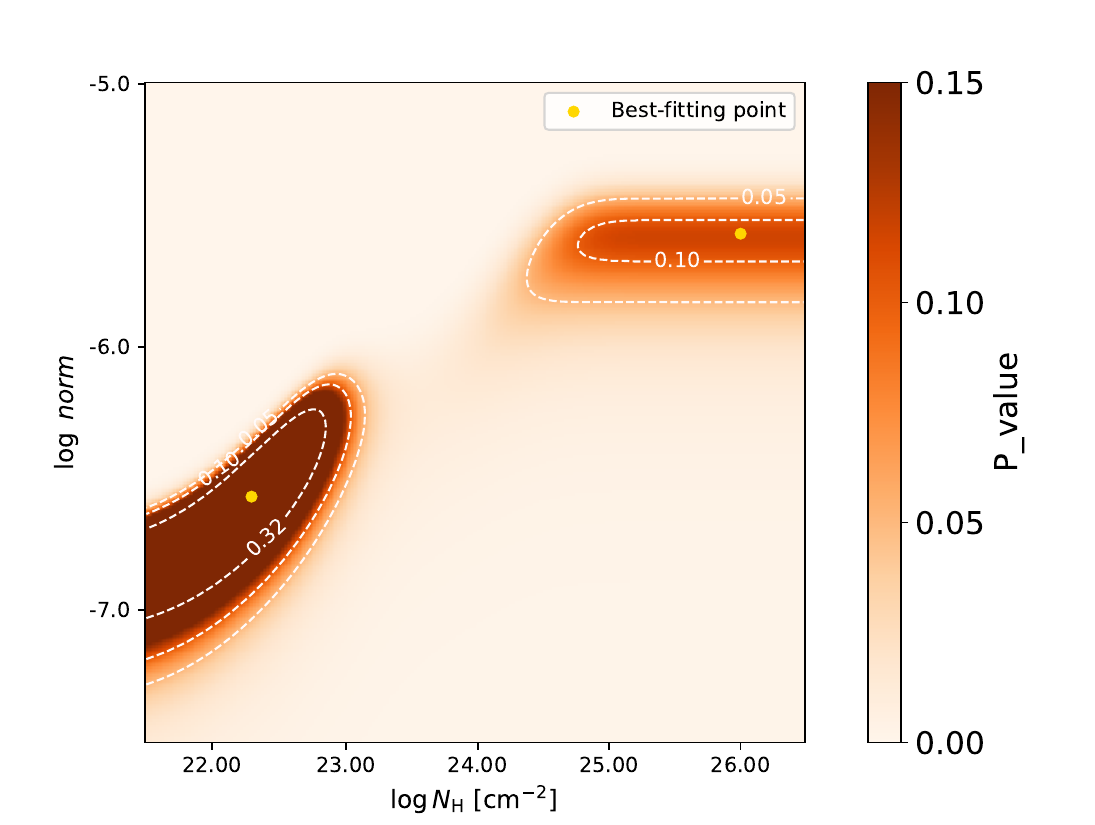}
        \caption{ p-value distribution in the $\log N_\mathrm{H}$–$\log$ \textit{\texttt{norm}} plane. White contours mark p = 0.05, 0.10 and 0.32.The golden solid circle indicates the best-fit point.}
        \label{fig:pvalue}
 \end{figure}

 To further constrain the absorption of these sources, we fitted the soft- and hard-band fluxes in \textit{Xspec} with the \texttt{\textit{zphabs$\ast$zpowerlw+constant$\ast$zpowerlw+pexmon}} model. The redshift was fixed at the mean value of the 23 sources (z = 0.92), the photon index at 1.9, the inclination angle  at 85,  and the scattered fraction at 0.005; only the column density, $N_\mathrm{H}$, and normalization \texttt{\textit{norm}} were allowed to vary, while all remaining parameters were kept at their default values.
Figure~\ref{fig:pvalue} presents the p-value distribution in the $\log N_\mathrm{H}$–$\log$ \textit{\texttt{norm}} plane obtained by fitting the soft- and hard-band fluxes with our model. 
In $\log N_\mathrm{H}$–$\log$ \textit{\texttt{norm}} plane, the data are consistent with the model at the 95\% confidence level within two distinct parameter regions, corresponding to Compton-thin and Compton-thick solutions, respectively. The golden solid circles mark the best-fitting points in both regions, with corresponding column densities of $2.09 \times 10^{22}\ \mathrm{cm}^{-2}$ and $9.55 \times 10^{25}\ \mathrm{cm}^{-2}$, respectively.
The X-ray upper limits of these 23 sources already fall below the relationships between MIR and X-ray luminosities after absorbing by Compton-thick material. If their absorption were located in the Compton-thin region, the majority of them would have to be intrinsically X-ray weak AGNs. However, the intrinsic incidence of X-ray weak AGNs is far lower than that of CT-AGNs, making it extremely improbable that most of them are X-ray weak AGNs. The Compton-thin solution can therefore be ruled out.
Within Compton-thick region, the 95\% confidence level demands $\log N_\mathrm{H} > 24.38$, while the 90\% confidence level requires $\log N_\mathrm{H} > 24.76$. Both thresholds lie well above the Compton-thick limit, providing direct and compelling evidence that these sources are CT-AGNs. 
As the hard X-ray band is only marginally detected (slightly above $1\sigma$) in the stacked spectrum, we conservatively treated its flux as an upper limit. Fitting the same model under this assumption still allows us to obtain a solution consistent with Compton-thick absorption.
 
\subsection{Missed diagnosis of CT-AGNs}

Although MIR-selected AGN samples are incomplete, numerous studies have shown that their CT-AGN fraction closely matches (and even has the potential to reach) the theoretical expectation of CXB \citep[e.g.,][]{2024ApJ...966..229L,2025ApJ...978..118B,2025MNRAS.540.3827A}.
Our MIR-selected sample excludes X-ray-detected AGNs, which preferentially select less absorbed AGNs. Consequently, we expect the CT-AGN fraction in our MIR-selected sample to markedly exceed that predicted by CXB models.
However, our MIR diagnostics identified only 23 CT-AGNs, 2.1\% (23/1104) of the MIR-selected sample, significantly below the fraction expected from CXB synthesis models.
This discrepancy clearly exceeds statistical error, indicating that MIR diagnostics miss a large number of CT-AGNs. Based on our analysis of the CT-AGN fraction within our sample, we infer that at least 300 CT-AGNs remain unidentified.
With the current data and methods, we are still unable to find these missing CT-AGNs. To identify these missed CT-AGNs, the most direct and efficient approach is to conduct significantly deeper X-ray exposures of the COSMOS in the future.
On the one hand, deeper exposures will enhance detection sensitivity, revealing AGNs that are currently buried in the background. This might not only confirm their existence, but also enable column density measurements via X-ray spectral fitting, thereby reliably identifying CT-AGNs.
On the other hand, for AGNs that remain undetected even after deep exposures, longer integrations will push their X-ray luminosity upper limits to lows, markedly enhancing the discriminatory power of MIR diagnostics.

\subsection{The properties of CT-AGN host galaxies}

CT-AGNs are widely regarded as an early, heavily obscured phase of AGN evolution \citep[e.g.,][]{2023PASP..135a4102G}. 
Their growth is intimately coupled to the host-galaxy environment, as evidenced by the tight M$_\mathrm{BH}$--$\sigma_\star$ relation \citep[e.g.,][]{2000ApJ...539L...9F} and the role of AGN feedback in regulating star formation \citep[e.g.,][]{2006MNRAS.370..645B}. Together, this picture supports a coevolutionary relationship between AGNs and their host galaxies \citep[e.g.,][]{2013ARA&A..51..511K}.
CT-AGNs are considered to preferentially host in galaxies with a significant amount of dense gas \citep{2015ApJ...814..104K,2017MNRAS.468.1273R}. Intense AGN radiation compresses their surrounding dense gas, potentially triggering star formation activity within their host galaxies.
The host galaxies of CT-AGNs in the Local Universe exhibit a high level of star formation \citep[e.g.,][]{2006ApJS..163....1H,2012ApJ...755....5G}.
\cite{2025ApJ...987...46Z} find that the host galaxies of CT-AGNs in the CDFS exhibit higher levels of star formation activity than those of non-CT-AGNs.
\cite{2022MNRAS.517.2577A} suggested that obscured AGNs have star formation rates that are about three times higher than unobscured systems in the COSMOS.
However, \cite{2022A&A...666A..17G} suggested that the host galaxies of high-redshift AGNs are different from those of AGNs in the Local Universe. 
\cite{2025A&A...694A.241G} reported similar stellar mass and SFR distributions for CT-AGN and non-CT-AGN host galaxies; while their sample contains only 18 CT-AGNs, which is insufficient for statistically robust conclusions. Using a substantially larger sample, we will re-examine the SFRs and stellar masses of CT-AGN host galaxies relative to those of non-CT-AGNs.

\begin{figure}
        \centering
        \includegraphics[width=1.0\linewidth]{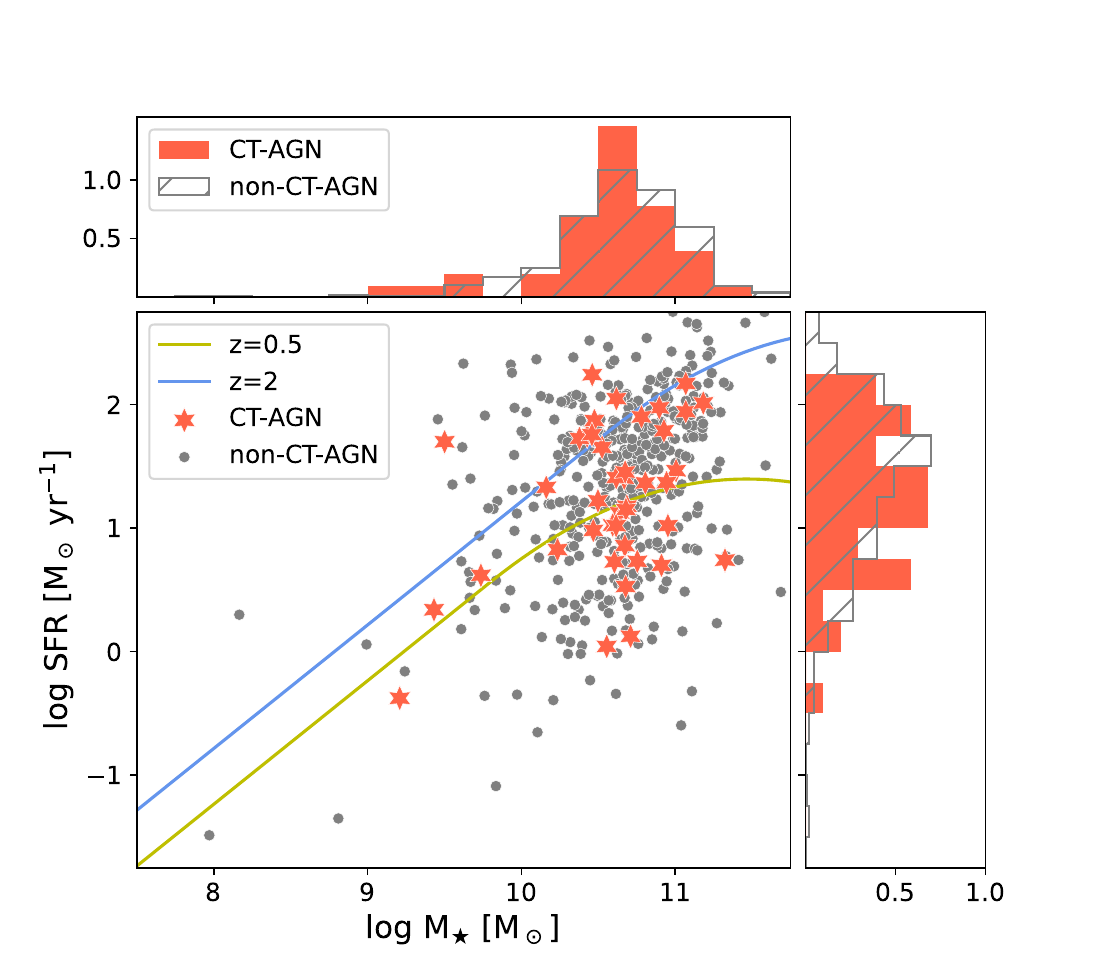}
        \caption{Relationship of the main sequence in AGN host galaxies. The solid red stars indicate the CT-AGNs and the gray circles represent non-CT-AGNs.}
        \label{fig:SFR-M}
\end{figure}

Because a significant number of CT-AGNs in our sample remain unidentified, we were not able to extract a pure non-CT-AGN sample. We therefore adopted the non-CT-AGNs provided by \cite{2025A&A...694A.241G} as the control sample and then combined their CT-AGNs with the 23 newly identified in this work to construct the final CT-AGN sample.
Figure~\ref{fig:SFR-M} presents relationships, represented by the main sequence, between SFRs and stellar masses of the host galaxies for both CT-AGNs and non-CT-AGNs.
A Kolmogorov–Smirnov test (KS test) of stellar mass distributions for their host galaxies gave us $p=0.33$, while repeating the KS test on their SFR distributions yielded $p=0.20$. Although these values are slightly lower than those reported by \cite{2025A&A...694A.241G}, they still indicate that there is no significant difference between the two distributions.
Thus, our findings do not support any correlation between intense star formation and CT-AGNs, as reported by  \cite{2025ApJ...987...46Z}. This finding is likely due to the limited size of our CT-AGN sample, which is still too small for statistically robust conclusions. In future works, we will enlarge the CT-AGN sample and further study whether there are differences in the host galaxies of both CT-AGNs and non-CT-AGNs.

\section{Summary}\label{sec:summary}
In this work, we aim to identify CT-AGNs hidden in AGNs whose X-ray flux is below the current flux limits in the COSMOS.
First, we selected a sample of 1,104 MIR-selected AGNs that were covered, but individually undetected in X-ray. 
Then, we reduced the \textit{Chandra} observational data covering the field and derived upper limits on the X-ray luminosity for all AGNs in our sample.
We subsequently collected multiwavelength photometric data for the sample and extracted key physical parameters for each source through SED fitting, such as six~$\mu$m~luminosities for AGNs, stellar masses, and SFRs.  
In the next step, using an MIR diagnosis, 7 to 23 sources could be identified as CT-AGNs. To confirm that our CT-AGNs are indeed heavily absorbed in the X-ray band, we performed a stacking analysis. 
The stacked signal was detected at 3$\sigma$ significance in the soft band, with a flux of \((7.57\pm 2.25) \times10^{-17}\ \mathrm{erg\ cm^{-2}\ s^{-1}}\) , whereas the hard-band excess is only 1$\sigma$ significant, corresponding to \((2.34\pm 1.55)\times10^{-16}\ \mathrm{erg\ cm^{-2}\ s^{-1}}\).
Fitting the stacked soft- and hard-band fluxes with the \texttt{\textit{zphabs$\ast$zpowerlw+constant$\ast$zpowerlw+pexmon}} model allowed us to obtain two parameter regions that are consistent with the data at the 95\% confidence level, corresponding to Compton-thin and Compton-thick solutions, respectively. 
The Compton-thin solution can be ruled out, leaving the Compton-thick absorber as the only viable scenario and confirming these sources as CT-AGNs.
We fit their soft- and hard-band fluxes with the \texttt{\textit{zphabs$\ast$zpowerlw+constant$\ast$zpowerlw+pexmon}} model and found that when Compton-thick absorption was included,  the model did provide a satisfactory fit to both bands. This provides direct and compelling evidence that these sources are CT-AGNs.
Although we have already identified some CT-AGNs in the MIR-selected AGNs, a considerable population of CT-AGNs was still missed by our selection.
Therefore, we have estimated and discussed the number of missed CT-AGNs in our sample and proposed a practical strategy to search for them.
Finally, based on our comparison of host-galaxy properties between CT-AGNs and non-CT-AGNs, we did not find any evidence to support a correlation between intense star formation and CT-AGNs.

\section*{Data availability}
Table~\ref{Tab:summary} is only available in electronic form at the CDS via anonymous ftp to cdsarc.u-strasbg.fr (130.79.128.5) or via http://cdsweb.u-strasbg.fr/cgi-bin/qcat?J/A+A/.

\begin{acknowledgements}
        We sincerely thank the anonymous referee for useful suggestions.
      We acknowledge the support of the \emph{National Nature Science Foundation of China} (No. 12303017).
      This work is also supported by \emph{Anhui Provincial Natural Science Foundation} project number 2308085QA33. 
      QSGU is supported by the \emph{National Nature Science Foundation of China} (Nos. 12192222, 12192220, and 12121003).
      L.S.Y. is supported by the \emph{National Nature Science Foundation of China} (No. 12503011).
      F.L. is is supported by the \emph{National Nature Science Foundation of China} (No. 12403042).
      Y.Y.C. is grateful for funding for the training Program for talents in Xingdian, Yunnan Province (2081450001).
      X.L.Y. acknowledges the grant from the \emph{National Nature Science Foundation of China} (No. 12303012), \emph{Yunnan Fundamental Research Projects} (No. 202301AT070242).
      H.T.W is supported by the Hebei Natural Science Foundation of China (No. A2022408002).
\end{acknowledgements}

%

\bibliographystyle{aa} 
\bibliography{bibtex.bib} 

\begin{appendix}




\section{Parameters of MIR-selected AGNs}
This section presents a tabular summary of the MIR-selected AGN sample used in this study, the relevant parameters, and the identified CT-AGNs.

        \begin{sidewaystable*}
        \caption{Parameters adopted for CT-AGN selection from the MIR-selected AGN sample and summary of the identified CT-AGNs (extract).}\label{Tab:summary}
        \centering
        \resizebox{\hsize}{!}{
                \begin{tabular}{lcccc rcccc ccc}
                        \hline\hline             
                        ID & R.A. & Decl. & z &$B_{3''}$& \multicolumn{1}{c}{$T_{\mathrm{exp}}$} & $f_{\text{2--10}}^{\text{up-limit}}$&$\log$ L$_{2-10}$&$\log \nu \mathrm{L}_{\nu}$(6~$\mu$m)&$\alpha_\mathrm{AGN}$&$\log \mathrm{M_\star}$&$\log \mathrm{SFR}$& CT flag\\
                        &($^\circ$)&($^\circ$)&&&\multicolumn{1}{c}{($\mathrm{ks}$)}&($10^{-16}\ \mathrm{erg\ cm^{-2}\ s^{-1}}$)&($\mathrm{erg\ s^{-1}}$)&($\mathrm{erg\ s^{-1}}$)&&($\mathrm{M}_\odot$) &($\mathrm{M}_\odot \ \mathrm{yr}^{-1}$) & \\
                        (1)&(2)&(3)&(4)&(5)&\multicolumn{1}{c}{(6)}&(7)&(8)&(9)&(10)&(11)&(12)&(13)\\
                        \hline
                        571 & 149.6744 & 1.6249 & 0.76 & 3 & 88.30 & 44.23 & 42.947 & 42.864 & 1.000 & 8.822 & 0.171 & 0\\
                        2050 & 149.6823 & 2.2929 & 1.99 & 7 & 183.52 & 32.07 & 43.698 & 43.820 & 0.773 & 10.135 & 0.746 & 0\\
                        3175 & 149.8748 & 2.6273 & 1.99 & 8 & 181.39 & 35.00 & 43.735 & 43.721 & 1.000 & 9.897 & 0.940 & 0\\
                        4623 & 150.3852 & 1.7087 & 1.48 & 5 & 190.98 & 25.80 & 43.340 & 43.241 & 1.000 & 9.252 & 0.729 & 0\\
                        61366 & 150.2570 & 2.2118 & 1.75 & 0 & 189.12 & 10.90 & 43.117 & 45.571 & 1.000 & 10.927 & 1.788 & 1\\
                        69254 & 150.5989 & 2.1318 & 1.10 & 4 & 193.94 & 22.83 & 43.010 & 43.596 & 1.000 & 9.214 & 0.103 & 0\\
                        69426 & 150.4179 & 2.2177 & 0.57 & 2 & 182.79 & 18.35 & 42.294 & 43.825 & 1.000 & 10.605 & 0.727 & 1\\
                        70003 & 150.6131 & 2.4660 & 1.98 & 6 & 219.78 & 24.63 & 43.578 & 43.681 & 0.864 & 9.773 & 1.110 & 0\\
                        106842 & 150.6229 & 1.9051 & 1.67 & 3 & 231.31 & 16.89 & 43.264 & 43.790 & 1.000 & 8.477 & 0.068 & 0\\
                        107028 & 150.3769 & 2.7157 & 0.94 & 6 & 190.92 & 28.35 & 42.952 & 44.980 & 1.000 & 10.897 & 1.974 & 1\\
                        109199 & 150.1546 & 2.8030 & 0.60 & 1 & 91.08 & 30.27 & 42.554 & 43.291 & 1.000 & 10.054 & 1.049 & 0\\
                        133669 & 149.8801 & 2.3016 & 1.95 & 6 & 209.43 & 25.85 & 43.585 & 43.988 & 1.000 & 10.028 & 0.942 & 0\\
                        134768 & 149.5189 & 2.3015 & 1.41 & 7 & 69.26 & 84.98 & 43.810 & 43.595 & 0.999 & 9.986 & 1.278 & 0\\
                        135400 & 149.8988 & 1.8028 & 1.92 & 11 & 192.42 & 39.97 & 43.763 & 43.779 & 1.000 & 9.760 & 1.165 & 0\\
                        136577 & 150.3656 & 2.3868 & 1.93 & 8 & 215.15 & 29.51 & 43.635 & 43.813 & 1.000 & 9.340 & 1.300 & 0\\
                        \multicolumn{1}{c}{$\vdots$} & $\vdots$ &$\vdots$ &$\vdots$ &$\vdots$ &\multicolumn{1}{c}{$\vdots$} &$\vdots$ &$\vdots$ &$\vdots$ &$\vdots$ &$\vdots$ &$\vdots$ &$\vdots$ \\
                        898638 & 149.4213 & 2.0407 & 1.48& 1  & 44.37 & 62.13 & 43.717 & 45.127 & 1.000 & 10.606 & 1.862 & 0\\
                        899015 & 149.6794 & 2.2080 & 1.02& 6  & 186.81 & 28.98 & 43.044 & 44.834 & 1.000 & 10.461 & 2.242 & 1\\
                        899719 & 150.4337 & 2.7086 & 2.00& 4  & 191.12 & 23.16 & 43.559 & 44.042 & 1.000 & 9.131 & 0.139 & 0\\
                        927043 & 150.5996 & 2.0402 & 1.20& 5  & 159.05 & 30.98 & 43.227 & 43.546 & 1.000 & 8.527 & -0.310 & 0\\
                        927564 & 150.1823 & 1.7008 & 0.72& 4  & 185.35 & 23.89 & 42.625 & 44.780 & 1.000 & 10.617 & 1.410 & 1\\
                        927986 & 150.5286 & 2.8744 & 1.51& 2  & 111.99 & 29.96 & 43.423 & 43.747 & 1.000 & 9.103 & -0.263 & 0\\
                        932400 & 150.5801 & 2.8731 & 0.79& 2  & 103.83 & 32.31 & 42.855 & 42.805 & 1.000 & 8.421 & 0.188 & 0\\
                        932754 & 150.3066 & 2.1149 & 0.41& 13  & 235.73 & 36.30 & 42.285 & 43.790 & 1.000 & 10.678 & 0.527 & 1\\
                        933121 & 150.5423 & 2.0232 & 1.62& 4  & 183.94 & 24.07 & 43.390 & 43.421 & 1.000 & 10.037 & 0.708 & 0\\
                        946717 & 150.1100 & 1.9535 & 0.54& 7  & 213.00 & 27.63 & 42.425 & 43.978 & 1.000 & 10.675 & 1.454 & 1\\
                        964368 & 150.1913 & 1.7899 & 0.70& 8  & 188.73 & 33.64 & 42.747 & 43.747 & 1.000 & 10.276 & 0.752 & 0\\
                        \hline
                \end{tabular}
        }
        \tablefoot{This table is available in its entirety in the machine-readable format. Only a portion of this table is shown here to demonstrate its form and content. Column (1) contains IDs from the COSMOS2020 FARMER catalog. Columns (2) and (3) contain the R.A. and decl. of the sources. Column (4) contains the photometric redshift that we used. Column (5) is the background counts of the source within a radius of $3''$. Column (6) represents the exposure time at the source location. Column (7) contains the upper limit of 2--10~keV flux. Column (8) contains the logarithm of 2--10~keV luminosity upper limit in the rest-frame. Column (9) contains the logarithm of 6~$\mu$m~luminosity for an AGN in the rest-frame. Column (10) contains the confidence of AGN components for SED fitting. Columns (11)--(12) contain the logarithm of M$_\star$, and SFR for AGN host galaxy. Columns (13): CT-AGN flag; "1" = CT-AGN identified in this work.}
\end{sidewaystable*}

\clearpage

\end{appendix}

\end{document}